\documentstyle[12pt,aasms4]{article}
\slugcomment{submitted to ApJ Lett.}

\def\lax    {\ifmmode{_<\atop^{\sim}}\else{${_<\atop^{\sim}}$}\fi}
\def\gax    {\ifmmode{_>\atop^{\sim}}\else{${_>\atop^{\sim}}$}\fi}

\def\gtorder{\mathrel{\raise.3ex\hbox{$>$}\mkern-14mu
	     \lower0.6ex\hbox{$\sim$}}}
\def\ltorder{\mathrel{\raise.3ex\hbox{$<$}\mkern-14mu
	     \lower0.6ex\hbox{$\sim$}}}
\def\gsim{\mathrel{\raise.3ex\hbox{$>$}\mkern-14mu
	     \lower0.6ex\hbox{$\sim$}}}
\def\lsim{\mathrel{\raise.3ex\hbox{$<$}\mkern-14mu
	     \lower0.6ex\hbox{$\sim$}}}

\begin{document}

\title{Phase Difference and Coherence as Diagnostics of Accreting
Compact Sources}

\author{ Xin-Min Hua\altaffilmark{1}, Demosthenes Kazanas and Lev
Titarchuk\altaffilmark{2}} 
\affil{LHEA, NASA/GSFC Code 661, Greenbelt, MD 20771}

\altaffiltext{1}{NRC/NAS Senior Research Associate}
\altaffiltext{2}{CSI, George Mason University}

%\vskip 0.5 truecm
%\font\rom=cmr10
%\centerline{\rom In press, Astrophys. J. (Letters)}

\begin{abstract}

We present calculations of the time lags and the coherence function of 
X-ray photons for a novel model of radiation emission from accretion
powered, high-energy sources. Our model involves only Comptonization 
of soft photons injected near the compact object in an extended but 
non-uniform atmosphere around the compact object. Our results show that 
this model produces time lags between the hard and soft bands of the 
X-ray spectrum which increase with Fourier period, in agreement with 
recent observations; it also produces a coherence function equal to 
one over a wide range of frequencies if the system parameters do not 
have significant changes, also in agreement with the limited existing 
observations. We explore various conditions that could affect coherence 
functions. We indicate that measurements of these statistical quantities 
could provide diagnostics of the radial structure of the density of this 
class of sources. 

\end{abstract}


\section{Introduction}

It is believed that the bright galactic X-ray sources are compact objects 
(black holes and neutron stars) powered by thermalizing the 
accretion kinetic energy on the surface of the neutron star or near the 
black hole horizon. The X-ray emission is, then, naturally accounted for 
as the result of Comptonization of soft photons by hot electrons which 
are expected to be present 
in the deep gravitational potential of a compact object. In fact, the 
spectra of these sources have been modeled successfully with this process, 
which has been analyzed in great depth theoretically (see e.g. Sunyaev 
\& Titarchuk 1980; Titarchuk 1994; Hua \& Titarchuk 1995) and much 
information about the conditions of the accreting gas can be derived from 
fitting the high energy spectra to observations.

It is well known, however, that the Comptonization spectra alone cannot 
provide any clues about the dynamics of accretion of the hot gas onto the 
compact object. One needs, in addition, time variability information. However, 
since it is tacitly accepted that the X-ray emission originates at the 
smallest radii of the accreting flow, one expects that such information 
should reflect the dynamical time scales or the electron scattering time 
scales associated with that region. In fact, the observed energy spectra 
indicate Thomson depths of a few and thus guarantee these two time scales 
to be of roughly the same order of magnitude. The recent RXTE observations 
(W. Focke, personal communication; also Meekins et al. 1984) which resolve 
shots of duration $\sim$ msec, appear to provide a validation for this 
simplest expectations. 

With these comments in mind, it appears strange that the X-ray fluctuation
power spectral densities (PSD) of accreting compact sources contain most of 
their power at frequencies $\omega \lsim 1$ Hz, far removed from the kHz 
frequencies expected on the basis of the arguments given above. This fact 
hints that one may have to modify the notion that the entire X-ray emission 
in this class of sources derives from a region a few Schwarzschild radii 
in size. Alternatively, the lack of high frequency power in the PSD 
could be attributed to the much longer viscous times scales of accretion 
disks near compact objects, or to an overall modulation of the accretion 
rate with an (otherwise undetermined) power spectrum similar to the 
observed PSD.

While the above arguments could provide reasoning for the observed
PSD form, they have hard time addressing observations associated with 
more involved tests of Comptonization induced variability and in 
particular the correlated variability at different energy bands. More 
specifically, Miyamoto et al. (1988, 1991) studied the time lags between 
the soft and hard photons in the X-ray light curves of Cyg X-1, using 
the GINGA data. It was shown in these references that the hard time lags 
increase roughly linearly with the Fourier period $P$ from $P \lsim 
$0.1 sec to $P \sim 10$ sec, being of order $0.01 P$ across this range 
of $P$. These long lags are very hard to understand in a model where 
the X-ray emission is due to soft photon Comptonization in the vicinity 
of the compact object. In such a model, they should simply reflect the 
photon scattering time in the specific region ($\simeq$ msec). The 
authors found the lag dependence on the Fourier period disturbing 
enough to question whether the process of Comptonization is indeed 
the one responsible for the formation of the X-ray spectrum of Cyg X-1.
In fact, this dependence also rules out scattering in an extended,
uniform, very low density X-ray corona, often invoked in models of 
accreting sources, since this too would produce time lags independent 
of the Fourier period.

More recently, Nowak \& Vaughan (1996) and Vaughan \& Nowak (1997,
hereafter VN97) have brought attention to another statistic of importance 
in understanding the spatio-temporal structure of accreting sources, 
namely the coherence of the X-ray light curves. This is to some extent 
the normalized cross-correlation function of the light curves at two 
different energy bands obtained from an ensemble of measurements. 
These authors computed the coherence function for the GINGA data of 
Cyg X-1 and GX 339-4 and found it to be equal to one for both sources 
over the frequency range 0.1 - 10 Hz. The coherence function for 
Cyg X-1 has also been computed with the more recent, higher quality 
data of RXTE (Cui et al. 1997b) and it was found to be equal to one 
up to frequency $\simeq$ 20 Hz. They consider this fact to be quite 
surprising, since most of the models they produce have coherence
functions substantially smaller than those obtained from observations.

Motivated by the discrepancy between the expected and the observed
variability behavior of accreting compact sources, Kazanas et al. (1997, 
hereafter KHT) proposed that the Comptonization process responsible for 
the formation of the spectra of accreting sources takes place in an 
non-uniform ``atmosphere" which extends over several decades in radius. 
It was shown that this model can account for the form of the observed 
PSDs, the energy spectra and at the same time predicts a correlation 
between the slopes of the PSD and the energy spectra.

In the present paper, we test this model further by studying its hard 
X-ray time lags and coherence function and examine under what conditions 
agreement with observation can be obtained. In \S 2 we briefly review the 
model of KHT and then compute the  associated time lags as a function of 
Fourier frequency. In \S 3 the coherence function is computed for the 
same model, while in \S 4 the results are reviewed and conclusions are 
drawn concerning possibility of uncovering the structure of accreting 
sources from spatio-temporal measurements.

\section{Time and Phase Differences}

The model of KHT considers Comptonization in a cloud of constant 
temperature but non-uniform density of profile $n(r) \propto r^{-1}$, 
which extends over several decades in the spherical radial coordinate 
from the compact source, $r$. In order to explore the Comptonization 
in clouds with density configurations such as these, we developed a 
Monte Carlo method which can treat photon propagation and Compton 
scattering in inhomogeneous media. The method was described in detail 
by Hua (1997) and its first calculations were displayed in KHT. 

The parameters of the calculations carried out in the present study 
were so chosen as to provide qualitative agreement of the resulting 
spectra with those of Cyg X-1 recently obtained by BATSE aboard CGRO. 
The CGRO/BATSE data, given by Ling et al. (1997), show that Cyg X-1 at 
its soft ($\gamma_0$) state has a spectrum consistent with 
Comptonization in an electron cloud of temperature $\sim 110$ keV and 
Thomson optical depth $\sim 0.45$. With this in mind, we employ a 
spherical model similar to that described in KHT but with temperature 
$kT_e = 100$ keV, in order to account for the excess emission at $\gsim 80$
keV. In fact, this spectrum extrapolates at low energies into the spectrum
of Cyg X-1 obtained by RXTE (Cui et al. 1997a) at a different epoch but 
with the source in a similar spectral state. The entire source has a 
radius $r_2 \approx 1.5$ light seconds and consists of a central core 
of radius $r_1$ and an extended ``atmosphere" with density profile 
$n(r) = n_1r_1/r$ for $r_1<r<r_2$. For the region $r < r_1$ we assume 
that it has a uniform density $n = n_1$, and that a soft photon source 
of blackbody spectrum at temperature 0.2 keV (Cui et al. 1997a) is 
located at its center (note that the density jump at $r=r_1$ used in KHT 
is absent in the present profile). 

As was shown in KHT, the density gradient of the extended atmosphere can 
significantly affect the resulting photon spectrum relative to that 
resulting from a uniform configuration of the same temperature and total 
Thomson depth. However, with a redefinition of the total Thomson depth, 
to account for the more efficient photon escape in this non-uniform 
configuration in comparison with the uniform one, the resulting spectra 
can be made similar. Thus, the total optical depth of the source in our 
investigation should be larger than the uniform source value ($\tau_0 = 
0.45$) used in Ling et al. (1997) to fit the BATSE data. We found that 
a total optical depth $\tau_0 = 1$ for our non-uniform configuration 
produces as good a fit to the BATSE data as the one used by Ling et al. 
(1997). Furthermore, we assume three values for the radius of the central 
core of the cloud: $r_1 = 2.4\times 10^{-2},~ 2.4\times 10^{-3}, ~2.4\times
10^{-4}$ light seconds. These conditions suffice to determine the density 
$n_1$, which is given by

$$n_1 = \displaystyle{\tau_0 \over {r_1\sigma_T \left[1+\ln 
\left(\displaystyle{r_2 \over r_1}\right)\right]}} ~~. \eqno(1)$$ 

With the above configuration, we have calculated, using the Monte Carlo code, 
the energies and arriving times of the photons emerging from the cloud under 
consideration to a distant observer. The photons are collected in the energy 
bands $2 - 6.5$ and $13.1 - 60$ keV in order to simulate precisely the 
recent RXTE observation (Cui et al. 1997b). In time, the photons are collected 
in 4096 bins, each with a width 6/4096 seconds. Based on the light curves so
obtained, we calculated the phase and time lags of the higher ($13.1 - 60$
keV) energy band with respect to lower one ($2 - 6.5$ keV) for clouds with 
the three different values of $r_1$ given above. The resulting time lags 
(solid curves) as well as the corresponding phase lags (dotted curves) as 
a function of the Fourier frequency are shown in Figure 1. It is apparent 
that these quantities have quite different form from those associated 
with Comptonization in uniform electron clouds (Hua \& Titarchuk 1996). 
For the latter, the phase lag functions have maxima at a characteristic
frequency roughly equal to the reciprocal mean escape time of the photons
from the cloud. The phase lags obtained from the present model, however, 
are almost constant, extending from a low frequency of $\simeq 0.1$ Hz, 
characteristic of the electron scattering time at the largest, least 
dense part of the atmosphere, to a cut off frequency at $\sim 100$ Hz, 
determined by the time resolution of our calculations. Consequently, the 
corresponding time lags are power-laws of indices $\lsim 1$, being at the 
lowest frequencies, as large as $\gtorder 0.1$ seconds, in rough agreement 
with observations of Cui et al. (1997b) and Miyamoto et al. (1988, 1991) 
(the slight increase in the lags at high frequencies is due to the effects 
of scattering in the uniform part of the configuration at $r < r_1$). 

It is apparent, that the relation of the time lags to the Fourier frequency
obtained in these simulations depends on the specific form of the radial 
dependence of density of the extended ``atmosphere". To indicate the effect 
of such a dependence, we also display in the same figure, the phase and 
time lags resulting from an atmosphere with density profile $n(r) \propto 
r^{-3/2}$, extending over a similar range in $r$. The difference between 
the curves corresponding to the different density profiles is evident. 
The lags associated with the steeper profile have a lot weaker dependence 
on the Fourier frequency, because most of the photon scatterings, which 
give rise to the lags, take place near the radius at which the Thomson 
depth is highest, i.e., at the smallest radii. Once the photon leaves
that region, it suffers very little additional scatterings and hence
little additional lag occurs at lower Fourier frequencies. In this sense, 
the density profile used in KHT and in the present study, is special in 
obtaining agreement with observation, a fact whose importance has not 
escaped the authors. Similarly, for density profiles flatter than that 
used herein, which cuts-off beyond a certain radius to ensure finite 
Thomson depth, most of the lags are produced by scattering at the largest 
radii and would therefore be representative of the scattering time at 
that radius. It becomes apparent therefore, that the observations of 
frequency dependent hard X-ray lags not only argues in favor of the 
presence of the extended atmosphere used herein, but also point to their 
study as a means for probing its detailed density profile. Furthermore, 
the range of the Fourier frequencies over which the phase lags are 
constant is indicative of the range in radii over which the specific 
power law profile of the atmosphere extends. We plan to provide a more 
extensive investigation of these issues in future publications.

\section {The Coherence Functions}

In search of more probing tests of the variability of accreting sources, 
VN97 have examined the coherence function. It basically indicates to what 
extent the light curves at two different energy bands track each other 
{\sl linearly} during the period over which the observations are made. 
These authors indicated that while most models of accreting compact 
sources have not been tested against this diagnostic, they generally 
tend to produce incoherent sources (Nowak 1994). Their conclusion was 
that it is very easy to destroy coherence and difficult to produce. 
Nevertheless, when analyzing GINGA data associated with Cyg X-1 and 
GX 339-4 they found (VN97) that the coherence function of these sources 
to be equal to one for Fourier frequencies below 10 Hz!

Motivated by these considerations we computed the coherence functions
for the model outlined in the previous section. 
In order to examine what conditions would lead to loss of coherence 
in our model, we assumed that the configuration our system changes 
during an observation. To simulate this evolution, we used the light 
curves in the energy bands $2-6.5$ and $13.1-60$ keV resulting from 
two configurations represented by different parameters of our model. 
In order to estimate the noise due to the statistical nature of the 
Monte Carlo calculations, we produced a group of eight light curves 
for each energy band of each configuration, obtained by following the 
history of $10^6$ photons with different initial random number seeds. 
The difference between each light curve and the average of the eight 
curves in the same group is taken as the noise. For the two energy 
bands given above, labeled 1 and 2 respectively, we used the 16 pairs 
of light curves, eight from each of two distinct configurations, and 
their respective noises to compute the coherence function defined in 
Equation (2) of VN97. The power spectra $\vert S_i\vert^2$ in this 
equation should be understood as being noise-corrected, that is, 
$\vert S_i\vert^2 = P_i -\vert N_i\vert^2 (i = 1,2)$, where $P_i$ and 
$\vert N_i\vert^2$ are power spectra obtained from the calculated light 
curve and the corresponding noise respectively. (We are indebted to B.
Vaughan for his insistence on this point.) The averages in the 
equation are taken over the 16 pairs of ``measurements".

For the purpose of verifying the above procedure, we first computed the 
coherence function of our model by averaging over 16 pair light curves 
from two identical configurations. We found that the coherence function 
of our model was 1 across the entire frequency range, as expected.
We then computed coherence functions for a variety of configuration 
evolutions with the results presented in Figure 2. It is seen that 
changes in the parameters of our model do result in loss of coherence 
over the frequency range $1/6 - 2048/6$ Hz. The four thick curves 
represent the evolution of the configuration starting from one of those 
described in the above section, namely that with $r_1=2.4 \times 
10^{-3}$ light seconds to one of the following configurations: (a) One 
with $n_1$ and $\tau_0$ increased by a factor of 5 with the 
density profile unchanged (solid curve). In this case, the physical 
sizes $r_1$ and $r_2$ remain the same and the coherence reduces to  
$< 0.6$ over the entire frequency range with large statistical 
fluctuations at high frequencies. (b) One with electron temperature 
decreased from the initial value of 100 keV to  50 keV (dotted curve). 
In this case, the coherence is $\approx 0.85$ below $\sim 100$ Hz 
and drops at higher frequencies. (c) Configurations with different energy
of initial soft photons. While in the initial configuration the source 
photons have a blackbody distribution at temperature $kT_0 = 0.2$ keV, 
the dashed curve is obtained between this initial configuration 
and one with source photons of a single energy $E_0 = 13.1$ keV. 
It is seen in this case the coherence is reduced virtually to zero. 
On the other hand, if the final configuration has source photons at 
the blackbody temperature $kT_0 = 4$ keV, the coherence becomes 
$\geq 0.8$ over the entire frequency range, as the dash-dotted curve 
indicates. (It has coherence slightly greater than 1 at high frequencies, 
probably due to noise overcorrection.)

Special attention should be paid to the thin curve in Figure 2. This
represents an evolution between two configurations described in the last
section, namely from that with $r_1 = 2.4 \times 10^{-2}$ light second
to that with $r_1 = 2.4 \times 10^{-4}$ light second (or vice versa).
It is seen that the coherence is virtually unity over the entire range 
of frequencies under consideration, although the size of the uniform part
of the configuration changes by two orders of magnitude. The cause of
this outcome becomes apparent by taking a closer look to the PSD and 
phase lag of these two configurations. Since the light curves are 
obtained from two different configurations, they essentially represent 
two independent time series, say $q$ and $r$ for the two energy bands 
1 and 2. One can then use Equation (10) in VN97 to understand their 
coherence. Using the same notation, we rewrite the equation in terms of 
the ratios of PSD in the two energy bands 
$\alpha_1 = \vert Q_1\vert/\vert R_1\vert$ and 
$\alpha_2 =\vert Q_2\vert/\vert R_2\vert$. 

$$ \gamma_I^2 = \displaystyle{{1 + \alpha_1^2\alpha_2^2+2\alpha_1\alpha_2
\cos(\delta\theta_r-\delta\theta_q)}\over{(1 + \alpha_1^2)(1 + \alpha_2^2)}}.
\eqno(2)$$ 

\noindent
It is found that at low frequencies, the PSD in the two energy bands 
are in proportion $\alpha_1 \approx \alpha_2 \approx 1$. And from Figure 1, 
we see that the difference in the phase lag
between these energy bands is small $\delta\theta_r-\delta\theta_q 
\lsim 10^{\circ}$; equation (2) then suggests that the coherence 
should be $\gamma_I^2 \simeq 1$. At higher frequencies, however, we found 
$\alpha_1 \not= \alpha_2$ and $\delta\theta_r-\delta\theta_q$ could be
as great as $50^{\circ}$ (see Figure 1). We might then expect the coherence
to be less than one at these frequencies. But, in this frequency range, 
the values of $\alpha_1$ and $\alpha_2$ are found to be much smaller than 1, 
yielding $\gamma_I^2 \simeq 1$. 
Therefore the coherence at high frequencies is  only superficial,
since neither of the conditions outlined in VN97 are satisfied, and it
is caused by the small values of $\alpha_1$ and $\alpha_2$ compared to one. 
In a similar way, one can examine the true causes for the coherences or 
lack thereof as displayed in Figure 2. For example, when we examine the
case corresponding to the dashed  curve in Figure 2, it was found that the 
near zero coherence over the entire frequency range is due to the fact 
that $\alpha_2 \ll \alpha_1$, so that $\gamma_I^2 \approx 1/\alpha_1^2 
\approx 0.01$.

In view of these results, one should be cautious drawing strong conclusions
from measurements of coherence close to one, such as those of Cyg X-1 and 
GX339-4 given in VN97. That may indicate the constant state of the 
responsible mechanism over 
the observation time. On the other hand, it may also indicate changes 
in the system to which the coherence statistic is insensitive. 

\section{Discussion and Conclusions}

We have presented above two  statistics  associated with a model for 
the time variability of the X-ray radiation emitted by accreting compact 
objects. In particular: Using a Monte Carlo code, we have computed the 
time lags between soft and hard photons and also the coherence of the 
X-ray light curves between the same energy bands resulting from Compton 
scattering in constant temperature but non-uniform density ``atmosphere"
which extends over several decades in radius. In our simulations  we have 
assumed that the sole source of phase differences is due to Compton 
scattering of soft photons by the hot electrons of the atmosphere; the 
sole source of loss of coherence is the change of the system during 
observation. In addition, we assumed that the source of photons resides 
near the center of the configuration.

Our results indicate that:

1. The hard photons lag behind the soft ones by amounts which increase
with the variability Fourier period. This behavior is distinctly 
different from that of Compton scattering by an electron cloud
of uniform density and in agreement with observations in both  the GINGA 
and the RXTE data.

2. The hard photon phase and time lags as a function of the Fourier 
frequency depend on the density profile of the extended scattering 
atmosphere. The observed lags are consistent with a density profile 
$n(r) \propto r^{-1}$ used in this study, with the range of frequencies 
over which the phase lags remain constant reflecting the range of radii 
over which this density profile holds. Accurate measurements of the
lag dependence on frequency could be used in the deconvolution of 
the density profile of the atmosphere from observations.

3. Our model generally produces coherence close to one. This, in 
view of the results of VN97 on the coherence of Cyg X-1, would indicate
that the parameters of this system remain constant over very long 
time scales (hours). The coherence function between two energy 
bands, within our model, can be reduced to less than one by changing 
the macroscopic parameters of the Comptonization cloud and/or the energy 
of the source photons. However, the inverse of this statement is not true; 
coherence equal to one is, for practical purposes, only a sufficient 
condition for the parameters of the configuration remaining constant.

We believe that the simplicity and the ability of our model to 
correctly reproduce these variability statistics argues for the 
correctness of its basic premises, namely the presence of a non-uniform 
atmosphere and the effects of Compton scattering for determining the 
variability of accreting compact sources. We also believe that further 
scrutiny of this model by detail comparisons of all the information 
available, including the PSD and energy spectra, could determine the  
temperature and density structure of these accretion flows. 

We would like to thank W. Cui, W. Focke and J. Swank for useful
discussion and for communicating to us prepublication results. 
We also thank B. Vaughan for his careful reading of our manuscript and
his valuable suggestions for the implementation of our calculation of 
coherence function.

\clearpage

\centerline{\bf FIGURE CAPTIONS}

\figcaption{
The time (solid curves) and phase (dotted curves) lags  between energy
bands $13.1-60$ and $2-6.5$ keV resulting from
the extended atmosphere of temperature $kT_e=100$ keV and optical depth
$\tau_0 = 1$. Three cases of different sizes for the central core radius
$r_1$ are shown. The electron density has the form $n_1r_1/r$ for 
radius $r>r_1$ and $n_1$ for $r\leq r_1$. Also shown are the phase and time
lags (dashed-dotted and dashed curves respectively) for a density profile 
$n(r) = n_1 (r_1/r)^{3/2}$ with $r_1 = 2.4\times 10^{-4}$ light seconds.    }

\figcaption{
Various coherence functions between emissions in energy bands $2--6.5$ and 
$13.1-60$ keV from five pair of configurations of our model. (See text
for the specific definition of each curve.)} 
\end{document}